\documentclass[ amsmath,amssymb,aps,prb,reprint,superscriptaddress,showpacs]{revtex4-1}

\usepackage{graphicx}
\usepackage{dcolumn}
\usepackage{bm}
\usepackage{amsmath}

\usepackage{soul}

\usepackage{xcolor}

\begin{document}

\title{Two-step phase transitions in Fe(Se,Te)}

\author{D.A.~Chareev}
\affiliation{Korzhinskii Institute of Experimental Mineralogy RAS, Chernogolovka, 142432, Russia}
\affiliation{Ural Federal University, Ekaterinburg, 620002, Russia }
\affiliation{Dubna State University, Dubna, 141980, Russia}
\author{A.A.~Gippius}
\affiliation{Faculty of Physics, M.V. Lomonosov Moscow State University, Moscow, 119991, Russia}
\affiliation{P.N. Lebedev Physical Institute of the Russian Academy of Science, Moscow, 199991, Russia}
\author{Y.A.~Ovchenkov}
\email[]{ovtchenkov@mig.phys.msu.ru}
\affiliation{Faculty of Physics, M.V. Lomonosov Moscow State University, Moscow, 119991, Russia}
\affiliation{MIREA - Russian Technological University, Moscow, 119454, Russia }

\author{D.E.~Presnov}
\affiliation{Faculty of Physics, M.V. Lomonosov Moscow State University, Moscow, 119991, Russia}

\author{I.G.~Puzanova}
\affiliation{Korzhinskii Institute of Experimental Mineralogy RAS, Chernogolovka, 142432, Russia}
\affiliation{Dubna State University, Dubna, 141980, Russia}
\affiliation{National University of Science and Technology `MISiS', Moscow, 119049, Russia }
\author{ A.V.~Tkachev}
\affiliation{P.N. Lebedev Physical Institute of the Russian Academy of Science, Moscow, 199991, Russia}

\author{O.S.~Volkova}
\affiliation{Faculty of Physics, M.V. Lomonosov Moscow State University, Moscow, 119991, Russia}
\author{S.V.~Zhurenko}
\affiliation{P.N. Lebedev Physical Institute of the Russian Academy of Science, Moscow, 199991, Russia}
\author{A.N.~Vasiliev}
\affiliation{Faculty of Physics, M.V. Lomonosov Moscow State University, Moscow, 119991, Russia}


\date{\today}
%

\begin{abstract}

In the studied crystals of FeSe$_{0.7}$Te$ _ {0.3}$, a structural phase transition occurs in two stages. At higher temperatures, the electronic subsystem undergoes a rearrangement, leading to a significant increase in elastoresistance. $^{77}$Se NMR data show an abrupt change in the relaxation rate during this transition. The final transition occurs at a temperature several degrees below and is also accompanied by anomalies in the electronic properties. Thus, in the Fe(Se,Te) series, similarly to the behavior of pure FeSe under pressure, the type of transition changes and intermediate state appears before the structural transition is suppressed. This similarity between the corresponding phase diagrams is explained by the same deformation of the iron coordination environment in Fe(Se,Te) compounds and in FeSe under pressure. Our findings provide new and significant information on the phase diagram of Fe(Se,Te) compounds and in particular suggest the possible existence of a triple point near the quantum critical point.

\end{abstract}


\maketitle
\thispagestyle{empty}

\section{Introduction}

Studying phase diagrams of superconducting materials, identifying all possible ground states, and determining the phase boundaries between them is essential to understand how superconductivity occurs.
In iron-based superconductors (IBS) \cite{kamihara2008iron}, there are magnetic orderings and structural phase transitions that occur without magnetic ordering, which may correspond to orbital orderings. The competition between different types of ordering and their relationship to the emergence of superconductivity is still under discussion. There was great interest in the "preemptive" or two-step phase transitions that occur in $122$ and other series \cite{PhysRevLett.103.087001, PhysRevB.83.134522, doi:10.1073/pnas.1015572108, PhysRevB.85.024534}. These transitions occur when a magnetic state precedes a nonmagnetic nematic state that exists within a fairly narrow temperature range.

In this paper, we report the observation of the two-step phase transition in the quasi-binary composition FeSe$_{0.7}$Te$ _ {0.3}$. For IBS, quasibinary compounds of the $11$ series are the simplest. These compounds make it possible to study the electronic properties of the main structural element of the family in almost ideal structures. In this series, FeSe is of particular interest because it exhibits a crossover to high-temperature superconductivity when subjected to hydrostatic pressure\cite{Terashima2015,Miyoshi2014,Kothapalli2016}.

Progress in the synthesis of high-quality samples of Fe(Se,Te) compounds with a low tellurium content \cite{Zhang182,ovchenkov2019nematic,ovchenkov2020multiband,PhysRevB.100.224516, huang2022plot, hou2024bulk} made it possible to study in detail a new region of the phase diagram, where structural phase transitions are suppressed. The phase diagram for this range is surprisingly similar to that of the FeSe phase diagram under pressure, reproducing many of its details. \cite{ovchenkov2023crossover, ovchenkov2024peculiarities}. However, the layered structure of FeSe has a peculiarity in that deformation of the local iron environment under pressure occurs in the same direction as when selenium is substituted by tellurium. Thus, the similarity of these phase diagrams may mean that the symmetry of the local iron environment plays a dominant role in both cases. This finding is in good agreement with other studies on the role of the local iron environment in the properties of IBS \cite{lei2011phase}. Most importantly, this means that changes in the electronic properties of FeSe under pressure that lead to high-temperature superconductivity are probably also implemented in some Fe(Se,Te) compounds.

For the FeSe$_{0.7}$Te$ _ {0.3}$ crystals studied, there are two adjacent temperatures at which physical property anomalies are observed. The lower temperature, $T_{N1}$, is  approximately 35~K and apparently corresponds to a structural transition, because elastoresistance reaches a maximum. Between the superconducting transition and the $T_{N1}$ point, the resistance follows $T^{2}$. Above $T_{N1}$, there is another anomaly in the temperature dependence of the resistance at approximately 42~K, which we will refer to as $T_{N2}$. At $T_{N2}$, some significant changes occur in the electronic subsystem, causing a kink in the temperature dependence of the Hall constant. Below this temperature, the elastoresistance effect is significantly enhanced, which we explain by a change in the type of orbitals located at the Fermi level. Near $T_{N2}$ NMR studies reveal an abrupt change in the relaxation rate $1/T_{1}$, which has never been observed in other phase transitions in the $11$ series. This jump also indicates the reconstruction of the electronic subsystem at $T_{N2}$.

The two-step structural phase transition in Fe(Se,Te) compositions is another peculiarity that is also present in the phase diagram of FeSe under pressure. The two-step structural phase transition precedes the transition to high-temperature superconductivity in these phase diagrams. This can be a peculiarity of the corresponding quantum critical point (QCP) that deserves further investigation.

\section{Experiment}

The studied crystals of FeSe$_{0.7}$Te$ _ {0.3}$ were prepared using the AlCl$_{3}$/AlBr$_{3}$/KBr mixture in evacuated quartz ampoules in permanent gradient of temperature \cite{CrystEngComm12.1989, chareev2016general, chareev2016synthesis}. The quartz ampoules with the Fe$_{1.3}$Te$_{0.5}$Se$_{0.5}$ charge and maximum quantity of salt mixture were placed in a furnace so as to maintain their hot end at a temperature of  435~$^{\circ}$C and the cold end at a temperature of  383~$^{\circ}$C. The chalcogenide charge is gradually dissolved in the hot end of the ampoule and precipitates in the form of single crystals at the cold end. After being kept for 13 weeks in the furnace, platelike iron monochalcogenide crystals were found at the cold ends of the ampoules.

Magnetic DC susceptibility $\chi (T)=M(H)/H$ was measured using a Quantum Design MPMS SQUID in a field of $H=10$ kOe. Electrical measurements were done on cleaved crystal samples with contacts made by Pt sputtering using a mechanical mask. Heat capacity measurements were carried out on the Quantum Design Physical Property Measurement System. Elastoresistivity was measured using the AC transport option of the Quantum Design PPMS system with a multifunctional insert. The samples were glued to a commercial piezoelectric transducer and the strain gauges were located on the other side of the piezoelectric device \cite{chu2012}.

To conduct the $^{77}$Se ($S = 1/2$) NMR experiment, small crystals of the sample were ground and melted in paraffin to avoid shielding currents throughout the sample. All measurements were carried out in a constant magnetic field of 5.5028 T using the standard Hahn spin echo method on an upgraded Bruker MSL spectrometer \cite{zhurenko2021}. Because of the relatively small broadening of the NMR line in the entire studied temperature range, it was possible to excite it entirely at one frequency point. Therefore, the spectra were obtained as the Fourier transform of the second half of the spin echo, and relaxation measurements were carried out at the frequency of the NMR line maximum. The rate of nuclear spin-lattice relaxation $1/T_{1}$ was measured by saturation recovery and inversion recovery methods for low and high temperatures, respectively.

\section{Results}
\subsection{Macroscopic properties}
\begin{figure}[h]
\centering
  \includegraphics[scale=0.5]{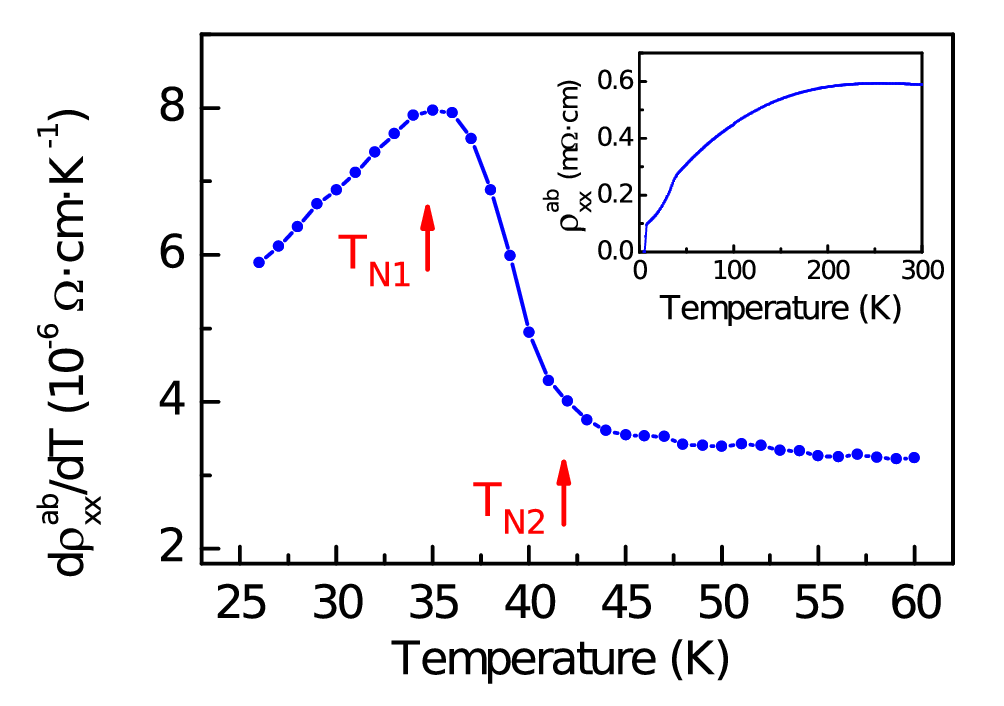}
  \caption{Temperature dependence of the derivative of resistivity ($d\rho_{xx}/dT$) in the $ab$-plane of crystals. Anomalies related to the phase transition are marked by $T_{N1}$ and $T_{N2}$. The inset shows the temperature dependence of $\rho_{xx}$ from room temperature down to helium temperatures. }
  \label{fgr:fig1}
\end{figure}

Magnetotransport properties, elastoresistance, magnetic susceptibility, and heat capacity were measured for the synthesized crystals of FeSe$_{0.7}$Te$ _ {0.3}$. Figure \ref{fgr:fig1} shows the temperature dependence of the longitudinal resistance measured in the $ab$ plane, $\rho_{xx}^{ab}(T)$ and its derivative. In the Fe (Se,Te) series of IBS, there is a change in the shape of the anomaly on the $\rho(T)$ curve at the point of the structural transition\cite{Zhang182,ovchenkov2019nematic,ovchenkov2020multiband,PhysRevB.100.224516}, indicating a change in the type or character of the transition. The compound we studied was from a region of the phase diagram where this change had occurred compared to unsubstituted FeSe.

For the studied FeSe$_{0.7}$Te$ _ {0.3}$, two breakpoints are distinguished on the derivative curve, designated as $T_{N1}$ ($\approx$35~K) and $T_{N2}$ ($\approx$42~K). Below $T_{N1}$, until the transition to the superconducting state, $\rho(T)$ follows a quadratic law and there are no signs of anomalies. This suggests that the main phase transition has been completed at $T_{N1}$.
However, changes in the behavior of $\rho(T)$, the Hall constant, and some other physical properties also occur at $T_{N2}$, which is approximately $6-7$ degrees higher. This allows us to discuss two stages of transition or a ``preemptive'' transition at $T_{N2}$.
In addition, for Fe(Se,Te) samples with increasing tellurium content, the evolution of the anomaly shape in the resistance temperature dependence at the phase transition point suggests that the point $T_{N2}$ originates from a structural transition point in the unsubstituted FeSe compound \cite{ovchenkov2023crossover}.

\begin{figure}[h]
\centering
  \includegraphics[scale=0.5]{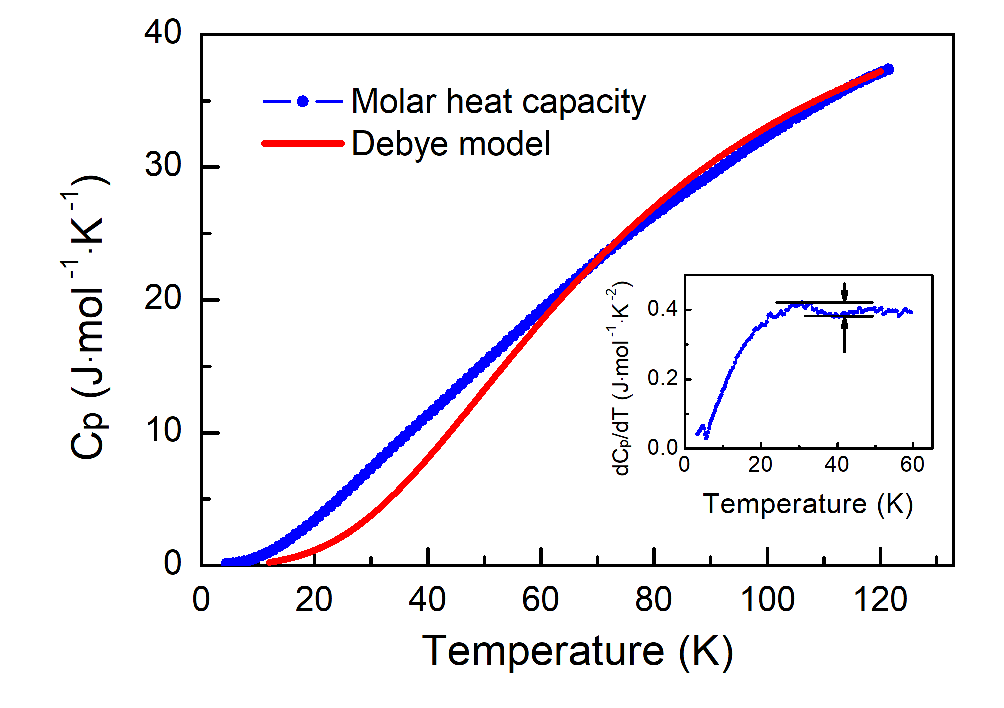}
  \caption{Molar heat capacity in the range from 2~K to 120~K and the Debye model for the Debye temperature equal to 300~K. The inset shows the temperature dependence of the derivative of the heat capacity. }
  \label{fgr:fig2}
\end{figure}

For FeSe, there is a clear step in the heat capacity during the transition from the tetragonal to the orthorhombic phase. For FeSe$_{0.7}$Te$ _ {0.3}$, the heat capacity curve $C_{p}(T)$ appears to be relatively smooth, as shown in Fig. \ref{fgr:fig2}. However, there is a peculiarity in the temperature dependence of the derivative of the heat capacity between $T_{N1}$ and $T_{N2}$, which can be interpreted as a downward departure from the expected smooth and convex behavior. During the structural transition in FeSe, the change in the heat capacity divided by temperature $\Delta{}(C_{p}/T)$ is 5.5 $mJ mol^{-1} K^{-2}$ \cite{PhysRevLett.114.027001}. For the composition studied, we can assume that $dC_{p}/dT$ decreases by 40 $mJ mol^{-1} K^{-2}$ in the temperature range between $T_{N1}$ and $T_{N2}$, so we can estimate the decrease in $C_{p}/T$ to be approximately 5-7 $mJ mol^{-1} K^{-2}$. Thus, heat capacity measurements suggest that there is a similar magnitude of entropy change during the phase transition in FeSe$_{0.7}$Te$ _ {0.3}$ as that observed during the structural transition in FeSe.

\begin{figure}[h]
\centering
  \includegraphics[scale=0.5]{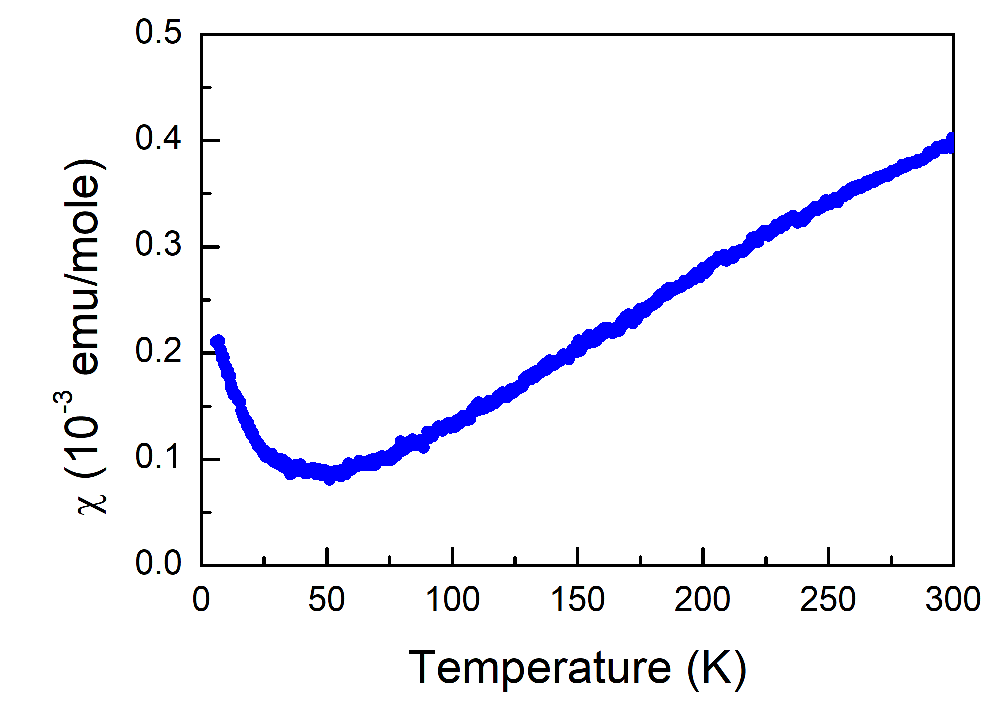}
  \caption{Temperature dependence of the DC magnetic susceptibility $\chi$.}
  \label{fgr:fig3}
\end{figure}

The temperature dependence of the magnetic susceptibility of FeSe$_{0.7}$Te$ _ {0.3}$ is shown in Fig. \ref{fgr:fig3}. Similarly to FeSe, the susceptibility increases slowly with temperature over a wide range and does not show any obvious anomalies during the structural phase transition. Unlike FeSe, there is a more significant increase in susceptibility at low temperatures, indicating the presence of magnetic moments. This could be a result of a change in the ground state and the presence of magnetic order in the sample, or it could be due to the presence of non-stoichiometric iron. Unfortunately, Fe(Se,Te) compositions are prone to stoichiometry violations, so the results of macroscopic magnetic measurements cannot be interpreted in favor of the microscopic magnetic order in FeSe$_{0.7}$Te$ _ {0.3}$.

\begin{figure}[h]
\centering
  \includegraphics[scale=0.5]{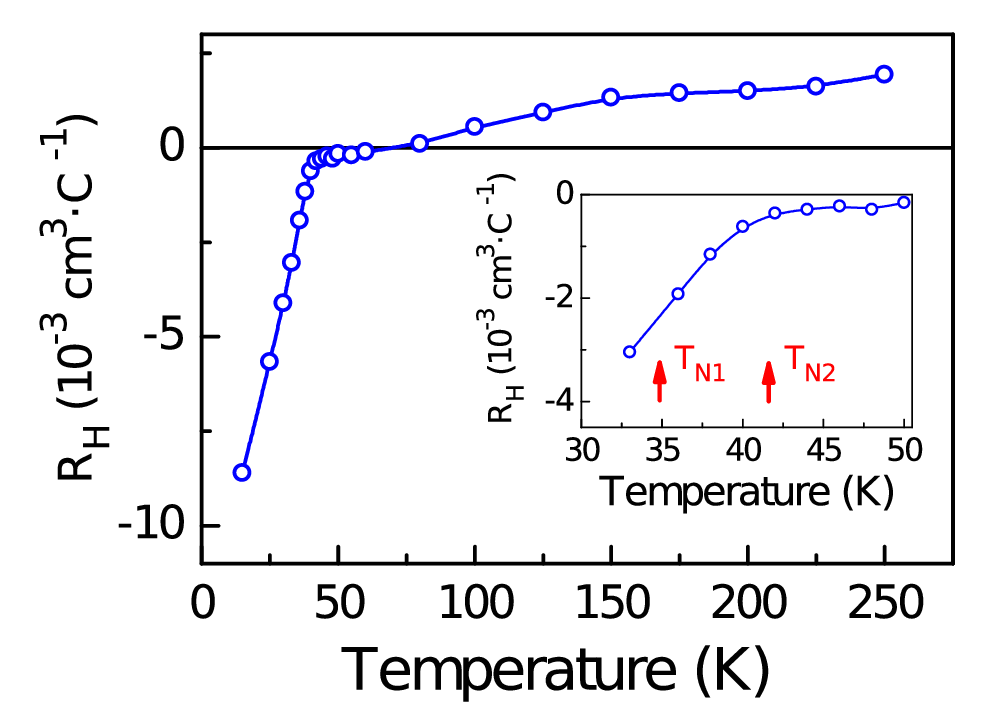}
  \caption{Temperature dependence of the Hall constant $R_{H}$. The inset shows a close-up view of this dependence around the phase transition region. $T_{N1}$ and $T_{N2}$ are points corresponding to anomalies in longitudinal resistance.}
  \label{fgr:fig4}
\end{figure}

The change in the details of the electronic subsystem reconstruction during the phase transition in FeSe$_{0.7}$Te$ _ {0.3}$ compared to FeSe can clearly be seen in the temperature dependence of the Hall constant $R_{H}$ plotted in Fig. \ref{fgr:fig4}. There is a kink in this dependence at $T_{N2}$, while for FeSe, the temperature dependence of the Hall constant remains smooth near the structural transition temperature. However, it should be noted that the field dependencies of $\rho_{xy}$ for FeSe are non-linear at temperatures below 50~K, making it difficult to unambiguously determine Hall constants. For FeSe$_{0.7}$Te$ _ {0.3}$, the dependencies $\rho_{xy}(B)$ are linear, although this may be a consequence of a general decrease in carrier mobility.

\begin{figure}[h]
\centering
  \includegraphics[scale=0.5]{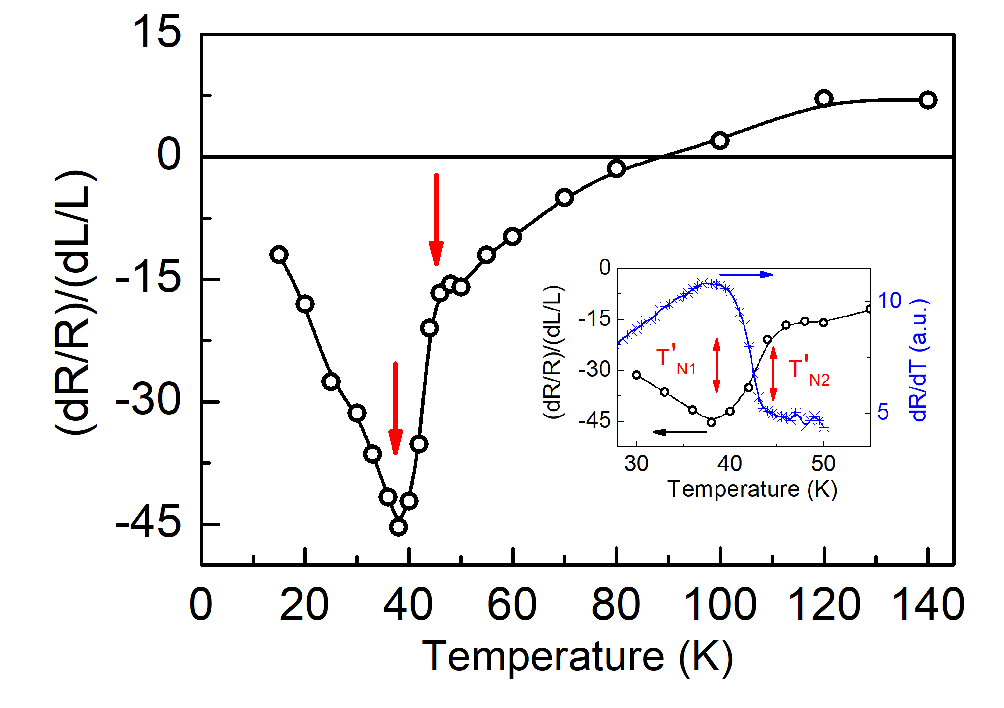}
  \caption{Temperature dependence of the elastoresistance ($dR/R$)/($dL/L$). Two arrows indicate anomalies related to the phase transition.The inset shows a close-up view of these anomalies along with the $dR/dT$ dependence measured on the same crystal which was glued to a piezostack for elastoresistance measurements. $T'_{N1}$$\approx$38 and $T'_{N2}$$\approx$42 corresponds to the same transition points as  $T_{N1}$ and $T_{N2}$ but shifted due to different measurements conditions. }
  \label{fgr:fig5}
\end{figure}

Changes in the electronic subsystem at $T_{N2}$ also affect the tensoresistive effect, as can be seen in Fig. \ref{fgr:fig5}. There is a kink in the temperature dependence of the elastoresistance ($dR/R$)/($dL/L$) at $T_{N2}$.
 FeSe, like other iron-based superconductors, exhibits a high elastoresistance effect. Optical investigation of IBS established the anisotropy of the Fermi surface parameters as the primary effect driving the $dc$ transport properties in the electronic nematic state \cite{mirri2015origin}. The elastoresistance of these compounds follows the Curie-Weiss law $C/(T-T_{NC}) + C_{0}$, where $C$, $C_{0}$, and $T_{NC}$ are some constants. The elastoresistance of FeSe reaches a maximum of around 40-50 units at the structural transition point. In a series of Fe(Se,Te) compounds with a low tellurium content, $C$ changes sign along with the change in the majority carrier type \cite{ovchenkov2020multiband}, similar to the tensoresistive effect in classical Si and Ge semiconductors. For FeSe$_{0.7}$Te$ _ {0.3}$ the sign of $C$ is negative. At $T_{N2}$, the rate of change in elastoresistance increases significantly and the magnitude of the effect within a narrow temperature range between $T_{N1}$ and $T_{N2}$ increases approximately three times. This increase can be attributed to the reconstruction of the Fermi surface. Below $T_{N2}$, the contribution to the electron density at the Fermi level from orbitals with a high degree of anisotropy in the tetragonal $ab$ plane is likely to be increasing. The subsequent maximum of elastoresistance at $T_{N1}$ indicates that a structural transition to lower symmetry occurs at this point.

\begin{figure}[h]
\centering
  \includegraphics[scale=0.5]{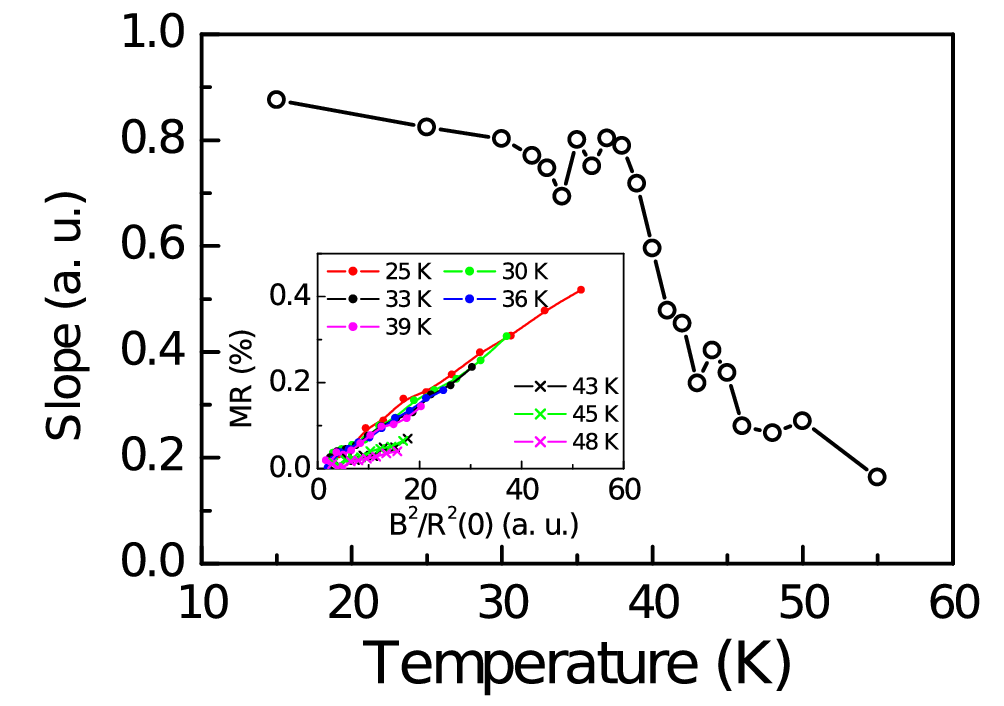}
  \caption{Temperature dependence of the slope of the linear approximation for the dependence of magnetoresistance $MR$=($R$(B)-$R$(0))/$R$(0) versus $(B/R(0))^{2}$. The inset shows the dependence of $MR$ versus $(B/R(0))^{2}$ (Kohler plot) for selected temperatures.}
  \label{fgr:fig6}
\end{figure}

Figure \ref{fgr:fig6} shows the results of the magnetoresistance measurements at different temperatures. These results demonstrate a violation of Kohler's rule for FeSe$_{0.7}$Te$ _ {0.3}$. Kohler’s rule states that magnetoresistance should be a function of the ratio of magnetic field $B$ to zero-field resistance $R(0)$. For the composition studied, the magnetoresistance is linearly dependent on the square of  $B/R(0)$, as shown in the insert of the figure. However, this slope is not constant with temperature. Kohler’s rule violation may indicate that Hall mobility is not equal to transport mobility, which is sometimes considered a sign of non-Fermi liquid behavior. For multiband semimetals, the violation may also be due to different temperature dependencies of the mobilities of the carriers or changes in their concentrations.

Kohler's rule is violated for FeSe. Nevertheless, the magnetoresistance of this compound can be satisfactorily described over a wide range of temperatures and magnetic fields using a quasi-classical approach that takes into account the presence of additional carrier groups \cite{ovchenkov2018mag}. In a two-band quasi-classical model, the slope of the $MR((B/R(0))^{2})$ curve depends on the ratio of the concentrations and mobilities of carriers.
Magnetoresistance can be expressed as follows:

\begin{eqnarray}
MR=\frac{\sigma_{1}\sigma_{2}(\mu_{1}-\mu_{2})^{2}B^2}{(\sigma_{1}+\sigma_{2})^{2}}\label{eq:1}
\end{eqnarray}

where $\sigma_{}=q_{i}n_{i}\mu_{i}$ is the conductivity of $i$-th band and $q_{i}$, $n_{i}$, and $\mu_{i}$ are the charge, the density, and the mobility of carriers with the convention that $q_{i}$ and $\mu_{i}$ carry the same sing. For compensated materials, the carrier concentrations coincide $n_{i}\equiv{}n$ and the slope of $MR((B/R(0))^{2})$ can be expressed as

\begin{eqnarray}
\frac{MR}{(B/R(0))^{2}}=\frac{-\mu_{1}/\mu_{2}}{n^{2}(1-\mu_{1}/\mu_{2})^2}\label{eq:2}
\end{eqnarray}

 Figure \ref{fgr:fig6} shows that, below the structural transition temperature, this slope remains almost constant, and Kohler's rule is satisfied.  Furthermore, it can be argued that Kohler's law is significantly violated near $T_{N2}$ because at $T$=39 K, which is above $T_{N1}$, the slope is approximately the same as at lower temperatures. Thus, the analysis of Kohler's rule also leads to the conclusion of a sharp change in carrier properties near $T_{N2}$.

Equations (\ref{eq:1}) and (\ref{eq:2}) allow us to consider possible reasons for violating Kohler's rule. In some cases, the "Extended Kohler's Rule"  is used, which takes into account changes in concentration \cite{PhysRevX.11.041029}. However, in order to explain our experimental results, the carrier concentration must be reduced by more than half during the transition to the low-temperature phase. This contradicts the notion that the transition occurs between the high-temperature state of a bad metal and the low-temperature state of a good metal.

The mobility ratio $\mu_{1}/\mu_{2}$ for FeSe is close to $-1$. A change in this ratio can generally explain the observed violations of Kohler's rule. At the same time, a possible change in this ratio is in agreement with the transition from bad to good metal and the emergence of an orbital-selective Mott phase in the Fe(Se,Te) series \cite{huang2022plot}. However, a change in the ratio of electron and hole concentrations can also make a comparable or even larger contribution. For example, if we put $\mu_{1}=-\mu_{2}\equiv{}\mu$ in equation (\ref{eq:1}), then we get 

\begin{eqnarray}
\frac{MR}{(B/R(0))^{2}}=\frac{2n_{1}/n_{2}}{(n_{1}+n_{2})^{2}(1+n_{1}/n_{2})^2}\label{eq:3}
\end{eqnarray}   

In our opinion, both mechanisms can play a significant role in violating Kohler's rule near the structural transition of the composition studied.

\subsection{$^{77}Se$ NMR}

\begin{figure}[h]
\centering
  \includegraphics[scale=1]{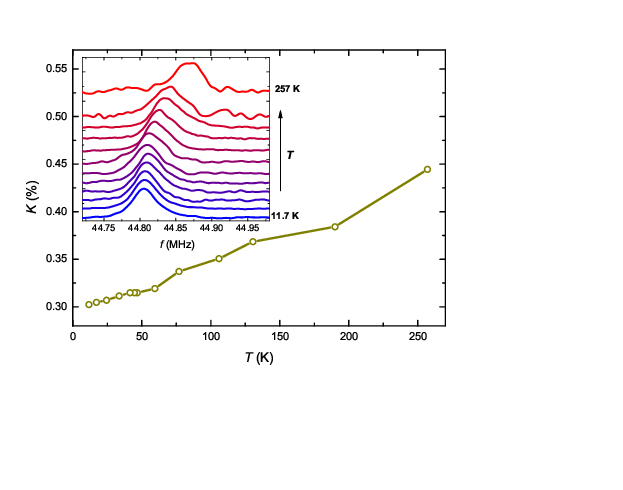}
  \caption{Temperature dependence of the $^{77}$Se NMR line isotropic shift $^{77}K$. Inset: frequency-sweep spectra at various temperatures.}
  \label{fgr:fig8}
\end{figure}

As mentioned in the Experimental section, the $^{77}$Se NMR spectra have the shape of a relatively narrow line with a full width at half maximum of about 40-50 kHz over the entire temperature range studied (see the inset in Fig. \ref{fgr:fig8}).  Its isotropic shift $^{77}K$ gradually decreases from 0.45\% at 257~K to 0.30\% at 11~K without any pronounced features. These values are typical for FeSe-based compounds, although some difference in the temperature dependence can be noted: no low temperature flattering of the $^{77}K(T)$ dependence can be observed, as in the parent FeSe \cite{PhysRevLett.102.177005,Bohmer2015PRL,baek2015orbital,PhysRevB.96.094528} and in some substituted compounds \cite{baek2020separate,PhysRevB.107.134507}. On the other hand, such behavior agrees well with the $\chi(T)$ dependence for the studied compound (Fig. \ref{fgr:fig3}), which has an almost constant slope, except for the Curie tail at the lowest temperatures, in contrast to the FeSe susceptibility, which typically exhibits a low temperature plateau (see, e.g., \cite{Bohmer2015PRL}). It is also in line with the NMR data for Te-rich compounds FeSe$_{1-x}$Te$_{x}$ $x \ge 0.5$, showing a uniform increase in $^{77}K$ with temperature starting from the superconducting transition for both $^{77}$Se and $^{125}$Te \cite{Arcon2010PRB,hara2011se,PhysRevB.82.064506,shimizu2009pressure}.

\begin{figure}[h]
\centering
  \includegraphics[scale=1]{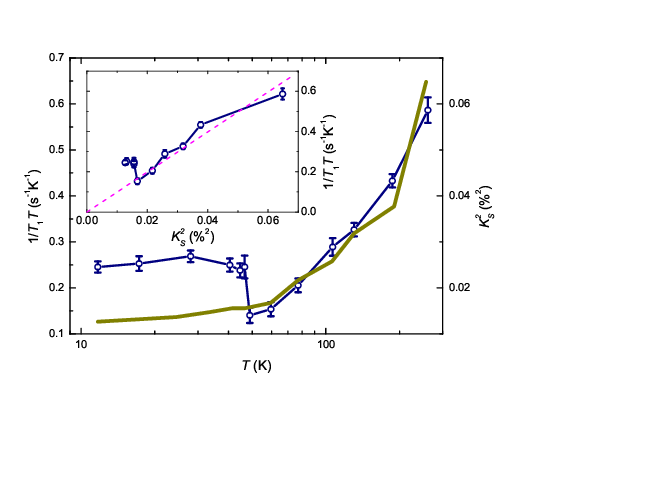}
  \caption{Temperature dependence of the reduced nuclear spin-lattice relaxation rate $1/T_1T$ (blue circles, left scale) and the squared spin part of the NMR isotropic shift $K_S^2$ (thick dark yellow line, right scale). Inset: $1/T_1T$ vs $K_S^2$ plot.}
  \label{fgr:fig9}
\end{figure}

The measured nuclear spin-lattice relaxation (NSLR) curves represent a simple single-exponential dependence over the entire temperature range studied. The NSLR rate $1/T_1$ increases linearly with temperature up to $\sim$45~K, providing an almost constant product of $1/T_1T$ (Fig. \ref{fgr:fig9}) in agreement with the Korringa’s law for relaxation via conduction electrons \cite{slichter2013principles}:

$$1/(T_1 T)=\pi \hbar^3 \gamma_e^2 \gamma_n^2 A_{hf}^2 N^2 (E_F ) k_B=SK_S^2 \biggl({\gamma_n\over\gamma_e}\biggr)^2  {4\pi k_B\over \hbar}, \eqno(4)$$

where $\gamma_{e,n}$ are the electron and nuclear gyromagnetic ratios, $A_{hf}$ is the transferred hyperfine coupling, $N(E_F)$ is the density of conduction electrons at the Fermi level at the Se site, $K_S$ is the spin part of the isotropic shift $^{77}K$, $S$ is the so-called Korringa ratio, characterizing the deviation of the electron system from the model of a non-interacting Fermi gas. Near $T_{N2}$, an almost twofold drop in $1/T_1T$ occurs, reminiscent to the behavior of the previously studied FeSe$_{0.675}$Te$_{0.3}$S$_{0.025}$ \cite{ovchenkov2024peculiarities}. It can typically be explained by a change in the charge carriers density $N(E_F)$, but can also be associated with their redistribution between bands with different local density and/or different hyperfine coupling $A_{hf}$. The latter interpretation agrees well with the above discussed magnetoresistance change in this temperature range (Fig. \ref{fgr:fig6}), pointing to change in the ratio of mobilities or populations between bands. Anyway, we can state a significant rearrangement of the electronic structure near $T_{N2}$.

After the drop at $\approx T_{N2}$, a gradual increase in the reduced relaxation rate $1/T_1T$ value is observed. This is a common phenomenon for FeSe-based compounds, related to the gradual increase of the magnetic susceptibility $\chi$ and hence $^{77}K$ with temperature \cite{PhysRevB.107.134507,PhysRevB.101.180503,Bohmer2015PRL}. Indeed, taking $K_S$ = $^{77}K$ – 0.19\% we achieve a rather good correspondence between the $1/T_1T(T)$ and $K_S^2(T)$ dependencies in the temperature range of 50-250 K satisfying the Eq. (4) (compare the blue circles and the thick dark yellow line in Fig. \ref{fgr:fig9}). The temperature-independent part of the $^{77}K$ shift was chosen to be 0.19\% so that the ratio of the squares of the spin shifts $K_S^2$ for 50K and 250K was equal to the ratio of the corresponding values $1/T_1T$. The plot of $1/T_1T$ vs $K_S^2$ given in the inset of Fig. \ref{fgr:fig9} also clearly evidences for this explanation, demonstrating a linear dependence in the range from $\sim$ 50 K to $\sim$ 250 K (compare with the dashed magenta line).

\section{Discussion and Conclusion}

In FeSe, a marked change in volume under pressure results from the collapse of the van der Waals-bonded region that separates the Fe$_{2}$Se$_{2}$ planes \cite{medvedev2009electronic}. The calculations made in the recent work \cite{zajicek2024unveiling} show an increase in the height of the Fe$_{2}$Se$_{2}$ layer $h$ in FeSe under pressure that means similar deformation of the local environment of iron to that in Fe(Se,Te). 
From the results of these calculations and using Vegard's rule, it can be estimated roughly that the substitution of 1\% of selenium with tellurium in FeSe causes approximately the same change in the $h/a$ ratio of the Fe$_{2}$Se$_{2}$ layer hight to the lattice period as at a pressure of 1~kBar. A very similar value for the ratio can be obtained by comparing the details of the phase diagrams of Fe(Se,Te) and FeSe under pressure.
The main difference between these phase diagrams is that $T_{c}$ is lower for Fe(Se,Te) compared to FeSe under pressure. However, we believe that this may be due to structural disorder rather than fundamental differences in behavior. There are many other details that support the connection between the phase diagrams. One of the significant similarities between these phase diagrams is that FeSe undergoes a transition to a bad metal state under pressure \cite{reiss2024collapse} that resembles the transition to a bad metal in the Fe(Se,Te) series. We also consider the presence of two-step transitions in the vicinity of the region where $T_{c}$ increases to be an important feature of the phase diagrams that we are discussing. This could be a manifestation of structural instability, which is closely related to high-temperature superconductivity.

The two-step structural transition occurs in the composition of Fe$_{1+x}$Te over a narrow range of $x$ values \cite{mizuguchi2012evolution}, where there is competition between orthorhombic and monoclinic magnetic structures. The transition temperatures in FeSe$_{0.7}$Te$ _ {0.3}$ are very similar to those of Fe$_{1+x}$Te, suggesting a possible relationship between these transitions. However, no signs of magnetic order were found in the FeSe$_{0.7}$Te$ _ {0.3}$ samples studied.

For FeSe$_{0.7}$Te$ _ {0.3}$, a two-step transition is observed near the supposed QCP and can be directly related to it. For example, similar features near QCP in phase diagrams of metallic magnets are well known \cite{RevModPhys.88.025006}. These quantum features arise due to fluctuations associated with proximity to QCP. These fluctuations can also contribute to superconductivity. In confirmation, it is worth noting the proximity of the values of the triple point temperature and the maximum of $T_{c}$ for the FeSe phase diagram under pressure.

The deformation of the iron environment in FeSe under pressure and in Fe(Se,Te) compounds leads to the formation of an ideal tetrahedral environment. This should cause a degeneracy in the $t_{2g}$ multiplet. This allows us to conclude that this degeneracy creates conditions for the emergence of a two-step transition. Our experimental data provide information on the transitions and properties of the intermediate phase.

At $T_{N2}$ temperature, there is a significant change in the electronic subsystem, indicated by a significant violation of Kohler's rule. The slope of the curves on the Kohler graph (Fig. \ref{fgr:fig6}) changes almost four times, which can be explained by the assumption that the carrier concentration decreases approximately by half at temperatures below $T_{N2}$. However, an increase in the nuclear spin-lattice relaxation rate $1/T_1T$ contradicts this assumption. Therefore, there must be more complex changes in the electronic subsystem.
It should also be concluded that the low magnetoresistance in the tetragonal Fe(Se,Te) phase is caused not only by the low carrier mobility, which Kohler's law takes into account, but also by significant deviations from the unit ratio of the concentrations $n_{1}/n_{2}$ or mobilities $\mu_{1}/\mu_{2}$ of the carrier groups, as follows from expressions (\ref{eq:2}) and (\ref{eq:3}). This suggests that in the tetragonal phase of the compounds studied, the electronic states are not well balanced.

In the temperature range between $T_{N2}$ and $T_{N1}$, a high value of elastoresistance deserves special attention. This may indicate that the isotropy of the transport properties in the $ab$ plane is preserved below $T_{N2}$.  Furthermore, it suggests an increased contribution of the $xz$ and $yz$ orbitals to the transport properties, since the nematic nature of these compounds is typically associated with these orbitals. A change in the sign of the elastoresistance with respect to FeSe could indicate that the contribution of these states on the hole sheet of the Fermi surface dominates. The degeneracy of the states $xz$ and $yz$ below $T_{N2}$ suggests that the transition at this temperature is mainly due to the redistribution of states at the Fermi level between $xy$ and pairs of states $xz$ and $yz$.

\begin{figure}[h]
\centering
  \includegraphics[scale=0.5]{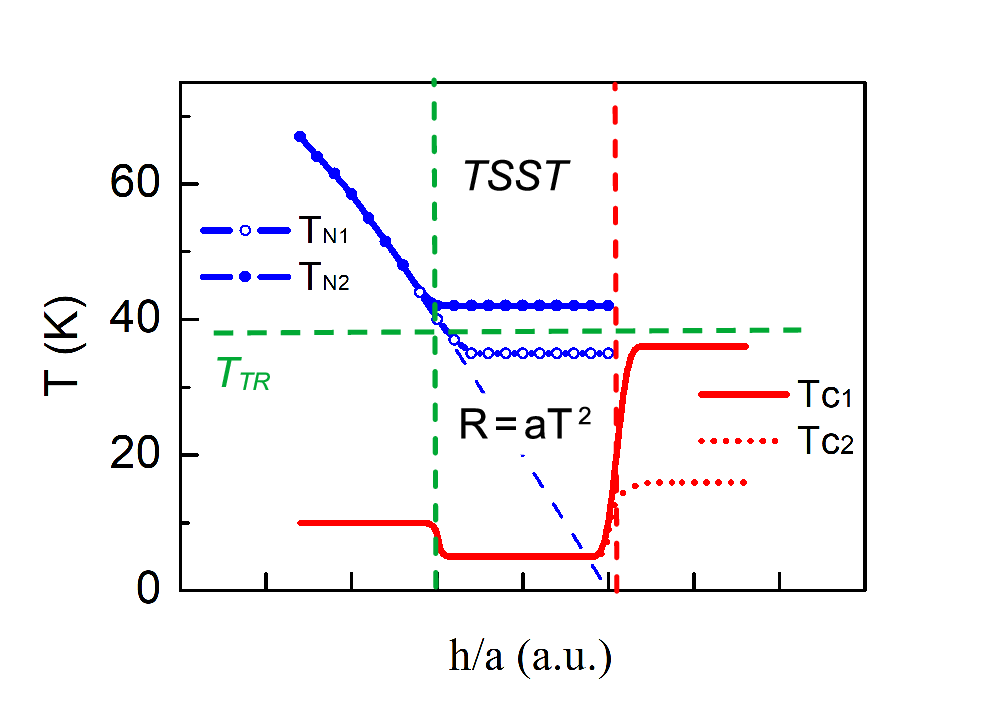}
  \caption{Supposed structure of the phase diagram for Fe(Se, Te) near the quantum critical point. The diagram shows the transition temperatures depending on the ratio of the Fe$_{2}$Se$_{2}$ layer hight $h$  to the lattice period $a$.  $T_{N2}$ and $T_{N1}$ are temperatures of structural transitions,  $T_{C1}$ is the critical temperature of the superconducting transition in FeSe under pressure, $T_{C2}$ is the critical temperature of the superconducting transition in Fe(Se,Te), which is possibly limited by structural disorder. The range where the structural transition occurs in two steps is labeled `TSST'. $T_{TR}$ stands for the temperature of a transition type change.   }
  \label{fgr:fig10}
\end{figure}

An extremely unusual for Fe(Se,Te), Fermi-liquid like metallic state is observed below $T_{N1}$, which may be the reason for the local $T_c$ minimum. What is unusual is that the resistance surprisingly strictly follows the T$^2$-law  in a wide range of low temperatures. This behavior is observed in a very narrow range of Fe(Se,Te) compositions \cite{ovchenkov2023crossover} near the QCP and the local minimum of $T_c$. On the basis of our new data, we suggest a possible structure of the phase diagram in the vicinity of the quantum critical point, as illustrated in Fig. \ref{fgr:fig10}).

In conclusion, it can be assumed that two-step transitions in the simplest binary series of iron-based superconductors reveal the peculiarity of the QCP, which may be related to the emergence of high-temperature superconductivity in this family. The corresponding QCP is connected to the local symmetry of the iron environment, and this implies that the details of the Fe(Se,Te) phase diagram deserve further study, both from a practical and fundamental research perspective.

\section{Acknowledgments}
This work was supported in part by Russian Science Foundation project 22-72-10034. Crystal growth funded by the Ministry of Science and Higher Education of the Russian Federation and Ural Federal University Program of Development within the Priority-2030 and FMUF-2022-0002. The equipment of the "Educational and Methodical Center of Lithography and Microscopy", M.V. Lomonosov Moscow State University was used.

\section*{References}
 \bibliographystyle{unsrt} 
\bibliography{FeSeTe_NCP.bib}

\end{document}